\documentclass[prb,a4paper,twocolumn,superscriptaddress]{revtex4}

\usepackage{amsmath,amssymb}

\newif\ifpdf
\ifx\pdfoutput\undefined
\pdffalse % we are not running PDFLaTeX
\else
\pdfoutput=1 % we are running PDFLaTeX
\pdftrue
\fi

\ifpdf
\usepackage[pdftex]{graphicx}
\else
\usepackage{graphicx}
\fi

\graphicspath{{./Figures/}}

\newcommand{\rmd}{\mathrm{d}} 
\newcommand{\RDW}{R_\mathrm{DW}} \newcommand{\ua}{\uparrow}
\newcommand{\da}{\downarrow} 
\newcommand{\ie}{\textit{i.e.\ }} \newcommand{\etal}{\textit{et al.\ }}
 \newcommand{\Amp}{\mathrm{A}}
 
\newcommand{\kohm}{\mathrm{k}\Omega} \newcommand{\eV}{\mathrm{eV}}
\newcommand{\metre}{\mathrm{m}} \newcommand{\micrometer}{\mu\mathrm{m}}
\newcommand{\nanometer}{\mathrm{nm}} 
 \newcommand{\second}{\mathrm{s}}

\newcommand{\LB}{Landauer-B\"uttiker }
\newcommand{\lwire}{L_\mathrm{wire}}
\newcommand{\stras}{\affiliation{Institut de Physique et Chimie des Mat\'eriaux
de Strasbourg, UMR 7504 (CNRS-ULP), 23 rue du Loess, Bo\^ite Postale 43, 67034
Strasbourg Cedex 2 FRANCE}} 
\newcommand{\uwa}{\affiliation{School of Physics, The University of Western
Australia, 35 Stirling Highway, Crawley WA 6009, AUSTRALIA}}
\renewcommand{\cos}[1]{\mathrm{cos}(#1)}
\renewcommand{\sin}[1]{\mathrm{sin}(#1)}

\newcommand{\sgn}[1]{\mathrm{sgn}(#1)}

\renewcommand{\eqref}[1]{Eq.~(\ref{#1})}

\begin{document}

\ifpdf 
\DeclareGraphicsExtensions{.pdf, .jpg, .tif} 
\else
\DeclareGraphicsExtensions{.eps, .jpg} 
\fi

\title{A Circuit Model for Domain Walls in Ferromagnetic Nanowires: Application
to Conductance and Spin Transfer Torques} 
\author{Peter E.\ Falloon}
\email{falloon@physics.uwa.edu.au} 
\uwa\stras 
\author{Rodolfo A.\ Jalabert} 
\stras
\author{Dietmar Weinmann} 
\stras 
\author{Robert L.\ Stamps} 
\uwa

\begin{abstract} 
We present a circuit model to describe the electron transport through a domain
wall in a ferromagnetic nanowire. The domain wall is treated as a coherent
4-terminal device with incoming and outgoing channels of spin up and down and
the spin-dependent scattering in the vicinity of the wall is modelled using
classical resistances. We derive the conductance of the circuit in terms of
general conductance parameters for a domain wall. We then calculate these
conductance parameters for the case of ballistic transport through the domain
wall, and obtain a simple formula for the domain wall magnetoresistance which
gives a result consistent with recent experiments. The spin transfer torque
exerted on a domain wall by a spin-polarized current is calculated using the
circuit model and an estimate of the speed of the resulting wall motion is made.
\end{abstract}

\maketitle

\section*{Introduction} 

Motivated by possible technological applications to non-volatile mass storage
devices, the subject of spin electronics has developed into a very active area
of research in recent years. Present-day state of the art devices are based on
the giant magnetoresistance (GMR) effect in ferromagnetic multilayer structures,
which was discovered in the late 1980's.\cite{baibich1988,binasch1989} Since
then, several exciting developments have led to new methods of storing and
switching the magnetic configuration of small magnetic elements. In particular,
the use of electric currents has been proposed as an alternative method to
reverse the magnetization of a magnetic layer or nanostructure. 

Within this context, attention has recently turned to the use of domain walls as
a possible basis for spin-electronic devices in ferromagnetic nanostructures.
Studies of magnetoresistance in ferromagnetic nanowires indicate that, in
addition to the well-known decrease in resistance due to the anisotropic
magnetoresistance effect, there is an increase in resistance during the
magnetization reversal process, which is attributed to the presence of domain
walls.\cite{ebels2000,dumpich2002} A domain wall trapped in a nanostructure can
thus provide a method for storing a bit of information, as has been demonstrated
in ferromagnetic nanocontacts \cite{garcia1999} and more recently in magnetic
semiconductor nanowires.\cite{ruester2003} Alternatively, domain wall
propagation under the influence of a spin-transfer torque induced by a
spin-polarized current can be used to reverse the magnetization of a nanowire,
providing a transport-based form of switching that does not require external
applied fields.\cite{grollier2002,vernier2004,yamaguchi2004}

Applications aside, spin transport through magnetic nanowires is also a
fundamental problem in mesoscopic physics, and a growing body of theoretical
work is appearing on the physics of electron transport through domain walls. In
the limit of narrow walls, calculations in the ballistic regime have shown that
reflection from the wall is the dominant source of
resistance.\cite{imamura2000,weinmann2001,gopar2004} This effect is large for
nanocontacts, but is very small when many transverse channels contribute, as in
the case of nanowires. On the other hand, calculations valid for wide walls have
shown that mistracking of electron spins traversing the wall results in an
enhancement of resistance similar to the GMR effect.\cite{levy1997,simanek2001}
However, the latter effect has not yet been considered for domain walls of
narrow or intermediate width, in which the spin mistracking becomes more
significant.

In this paper we introduce a circuit model for a domain wall which combines the
conductance properties intrinsic to the wall with the spin-dependent scattering
occurring in the region on either side. Our model is essentially a
generalization of the two-resistor model of Valet and Fert,\cite{valet1993}
which has been used to calculate the GMR effect in multilayers and interfaces.
The difference is that in place of an interface, which conserves the spin of the
current components, we have a domain wall, which mixes the two spin directions.
We mention also that our circuit model can be considered as a specific case of a
more general ``magnetoelectronic circuit theory'' for non-homogeneous mesoscopic
magnetic systems which has recently been
developed.\cite{brataas2000,brataas2001}

The layout of this paper is as follows. In section I we introduce our circuit
model and discuss the underlying physical assumptions, before deriving the
resistance of the circuit in terms of the intrinsic domain wall conductance
parameters in section II. In section III we derive these conductance parameters
for the case of a ballistic domain wall with small spin splitting of the
conduction band. We calculate the domain wall magnetoresistance within the
ballistic model and compare to recent experimental results in section IV. In
section V we consider the spin transfer torque exerted on the wall when a
current flows and obtain a simple estimate for the resulting velocity of the
wall. Finally, in section VI we discuss the limitations of our model and
consider possible extensions of our work.

%%%%%%%%%%%%%%%%%%%%%%%%%%%%%%%%%%%%%%%%%%%%%%%
\section{Physical description of circuit model}

In this section we discuss the physics underlying our circuit model. We start by
assuming a simple electronic band structure, the so-called $sd$ model, which
consists of an $s$ band of highly mobile free electrons, and a $d$ band of low
mobility electrons. The latter give rise to the magnetic domain structure of the
ferromagnet and contribute negligibly to the conductance. The $s$ electrons are
subject to the effective Hamiltonian
\begin{equation}\label{eq:hamiltonian} 
H = -\frac{\hbar^2}{2m}\nabla^2+
\frac{\Delta}{2}\vec{f}(\vec{r})\cdot\vec{\sigma},
\end{equation} 
where $\vec{\sigma}$ is the vector of Pauli spin matrices and $\vec{f}(\vec{r})$
is a unit vector representing the direction of the local magnetization due to
the $d$ electrons. The back-action of the $s$ electrons on the $d$ electrons is
neglected for the calculation of the transport properties. The energy $\Delta$
represents the strength of the spin-splitting of the $s$ band induced by its
interaction with the $d$ band. This splitting of the up and down spin sub-bands
results in a difference between the number of up and down states at the Fermi
energy. Here we use ``up'' to refer to the majority spin sub-band of the
incoming electrons, \ie the one with the lower potential (and hence larger
density of states at the Fermi energy $E_F$). 

Because of the lateral confinement in the nanowire, conduction electrons occupy
well-defined transverse modes. We assume for simplicity a wire with rectangular
cross-section $A=L_xL_y$, so that the modes are specified by quantum numbers
$n_x,n_y=1,2,\dots$ and energy 
\begin{equation} E_\perp = \frac{\hbar^2}{2m} \left[\left(\frac{\pi
n_x}{L_x}\right)^2 +\left(\frac{\pi n_y}{L_y}\right)^2\right]. \end{equation} 
The number of conducting channels for the up/down spin direction, $N_\pm$, is
equal to the total number of states at $E_F$ having longitudinal energy
$E_z=E_F-E_\perp$ in the range $-\sigma\Delta/2\le E_z\le E_F$. For wires of
physical interest, there is typically a large number of such states, and the
number of channels is approximately
$N_\sigma=(2mA/\pi\hbar^2)(E_F+\sigma\Delta/2)$. In this work we assume that the
relative magnitude of the spin splitting is small ($\Delta\ll E_F$) so that
$N_+\simeq N_-\simeq N=2mAE_F/\pi\hbar^2$. This assumption is valid for
ferromagnetic metals, where $\Delta/E_F\simeq0.01\textrm{--}0.1$, but not for
certain other systems, such as some magnetic semiconductors, where the
polarization can be as much as 100 percent.\cite{vignale2002}

In the ``two-resistor'' model of Valet and Fert \cite{valet1993} the key
physical assumption is that the length scale for scattering events which reverse
spin direction (the so-called spin diffusion length $l_{sd}$) is much larger
than the 
%inelastic mean free path which limits the 
phase coherence length $l_\phi$ (over which the orbital part of the wavefunction
loses coherence). Over length scales up to $l_{sd}$, the transport can thus be
modelled as two resistances in parallel, representing the two spin channels.
Since the resistivity is in general spin-dependent, this model can be used to
understand the GMR of an interface between two ferromagnetic layers, \ie the
difference between the resistance of the parallel and anti-parallel
configurations (both shown in Fig.~1). Letting $R_\pm$ denote the resistance of
the majority/minority spin channels over the length $l_{sd}$, the relative
increase in resistance of the anti-parallel configuration relative to the
homogeneous case is $(R_+-R_-)^2/2R_+R_-$. In the anti-parallel case the
currents in each channel are equal, while in the parallel case the total current
has a net polarization $\beta = (R_--R_+)/(R_++R_-)$.

\begin{figure}\label{fig:resistors}
\includegraphics[width=0.25\textwidth]{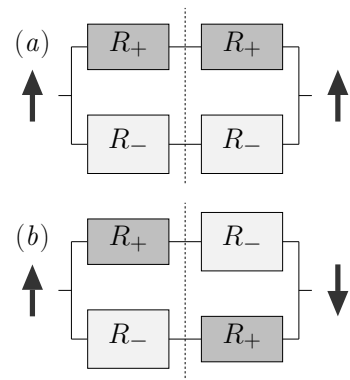}
\caption{The two-resistor model applied to an interface between two
ferromagnetic layers, in which the magnetization is (\textit{a}) parallel and
(\textit{b}) anti-parallel.}
\end{figure}

A domain wall is a region where the magnetization direction reverses over a
length which we denote $2\lambda$ (typically on the order of
$10\textrm{--100}\nanometer$). Electronic transport through this region is
characterized by a precessional motion in which the electron spins are partially
reversed as they track the rotating magnetization direction.\cite{viret1996} In
the so-called \emph{adiabatic} limit, $\lambda\rightarrow\infty$, this tracking
is perfect and the spins of transmitted electrons are completely reversed upon
traversing the wall. Incident majority (minority) electrons are transmitted into
the majority (minority) sub-band, and hence from the point of view of resistance
this limit is equivalent to a homogeneous magnetic configuration. At the other
extreme is the \emph{abrupt} limit, $\lambda\rightarrow0$, in which electrons
are transmitted with no spin reversal. In this case, incident majority
(minority) electrons are transmitted into the minority (majority) sub-band,
which corresponds to the antiparallel configuration of the above-mentioned
two-resistor model. The physically relevant regime for domain walls in nanowires
is generally intermediate between these two limits, with electrons tracking the
magnetization to varying degrees depending on their longitudinal
velocity.\cite{gopar2004} The ``mistracking'' of the electron spin with respect
to the wall magnetization results in a mixing of the up and down spin
directions. Such an intermediate case is beyond the scope of the circuits in
Fig.~1, and we are thus led to a modified circuit which includes the spin-mixing
behaviour of the domain wall. In our circuit model the latter is represented as
a 4-terminal circuit element, connecting incoming and outgoing currents of both
spin sub-bands.

Fig.~2 shows a sketch of our circuit model. The four-terminal element
representing the domain wall (DW) is connected to resistances $R_a$ representing
the diffusive spin-dependent transport occurring over a length $l_{sd}$ on
either side of the wall. With respect to a fixed quantization axis, electrons in
terminals 1 and 4 have spin $\hbar/2$, while those in terminals 2 and 3 have
spin $-\hbar/2$. $R_1$ and $R_4$ are equal to the majority resistance $R_+$,
while $R_2$ and $R_3$ are equal to the minority resistance $R_-$. A key feature
of the circuit is that the potentials in spin up and spin down channels close to
the wall, $V_1$ and $V_2$ ($V_3$ and $V_4$), are not necessarily equal. This
allows the distribution of current between the spin up and down channels to
differ from that of a homogeneous wire, giving rise to a GMR-like enhancement of
resistance. The spin-independent resistances $R_\mathrm{L}$ and $R_\mathrm{R}$
represent the resistance in the remainder of the wire, in which the two spin
channels are equilibrated. Experimentally fabricated nanowires typically have
lengths on the order of micrometers, and hence in practice $R_R,R_L\gg R_\pm$.

\begin{figure}\label{fig:circuit}
\includegraphics[width=0.45\textwidth]{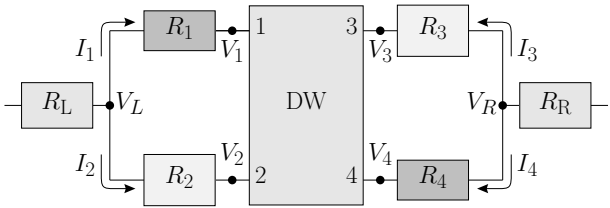}
\caption{The circuit model for a domain wall used in this paper.}
\end{figure}

Transport in each spin channel is treated classically in the two-resistor model,
since $l_\phi$ is assumed to be the smallest relevant length scale. However,
domain walls can have lengths on the order of $10\nanometer$ (in cobalt), or
even smaller in the presence of constrictions,\cite{bruno1999} which we might
expect to be comparable to $l_\phi$ at liquid nitrogen temperatures. It is
therefore necessary to adopt an approach based on phase coherent transport. To
see how the resistances $R_\pm$ arise in such an approach, we consider the
incoherent transport over the length $l_{sd}$ as a series of phase coherent
segments of length $l_\phi$. In each of these segments the transport is coherent
and diffusive, with a (spin-dependent) elastic mean free path $l_\pm$ and
resistance $R_\phi^\pm=\frac{h}{e^2} \frac{1}{N_\pm} \frac{l_\phi}{l_\pm}$. The
total number of phase coherent segments for each spin direction is
$\mathcal{N}_\pm=l_{sd}/l_\phi$, and hence the total resistance is $R_\pm =
\mathcal{N}_\pm R_\phi^\pm = \frac{h}{e^2}\frac{1}{N_\pm}\frac{l_{sd}}{l_\pm}$.
The spin-dependence of the resistance thus arises from differences in both
$N_\pm$ and $l_\pm$ between the two spin sub-bands. In our model it is assumed
that $N_+>N_-$, but both $l_+<l_-$ and $l_+>l_-$ are possible. In reality, which
of the latter pair of conditions is fulfilled is determined by the
material-dependent band structure, which governs the number of available states
at the Fermi energy into which electrons can be scattered. As mentioned
previously, we assume that the difference between $N_+$ and $N_-$ is rather
small, and hence in our model the difference between $l_+$ and $l_-$ is the
dominant factor in determining the spin-dependence of the resistance.

In the remainder of this paper we use the circuit of Fig.~2 to study two main
problems: domain wall magnetoresistance, in Sections II--IV, and current-driven
torques, in Section V.

\section{Resistance of domain wall circuit}

In this section we derive the resistance of the circuit in Fig.~2. For a general
domain wall, the currents and potentials at the wall can be related by the
multi-terminal \LB formula:\cite{buttiker1986,datta1997}
\begin{equation}\label{eq:lb conductance} 
I_{a} = \sum_{b \neq a} G_{ab}(V_{a} - V_{b}),\quad a,b = 1,2,3,4.  
\end{equation} 
The parameters $G_{ab}$ are matrix elements of the domain wall conductance
tensor for current passing between terminals $a$ and $b$. Since the Hamiltonian
\eqref{eq:hamiltonian} is symmetric under time-reversal (\ie there is no orbital
magnetic field term), these parameters satisfy $G_{ab} = G_{ba}$.\cite{datta1997}

\eqref{eq:lb conductance} can be written in matrix form as 
\begin{equation} \label{eq: G matrix}
\left(\begin{array}{c}I_{1}\\I_{2}\\I_{3}\\I_{4}\end{array}\right) =
\left(\begin{array}{rrrr} G_{11} & -G_{12} & -G_{13} & -G_{14} \\ 
-G_{21} & G_{22} & -G_{23} & -G_{24} \\ 
-G_{31} & -G_{32} & G_{33} & -G_{34} \\ 
-G_{41} & -G_{42} & -G_{43} & G_{44} 
\end{array}\right)
\left(\begin{array}{c}V_{1}\\V_{2}\\V_{3}\\V_{4}\end{array}\right),
\end{equation} 
where we have defined $G_{aa}=\sum_{b\neq a}G_{ab}$. By Kirchhoff's law,
the terminal voltages $V_a$ are related to the currents $I_b$ by
\begin{equation}
\begin{array}{cc} 
V_1 = V_L - I_1R_1, &\quad V_3 = V_R - I_3R_3, \\ 
V_2 = V_L - I_2R_2, &\quad V_4 = V_R - I_4R_4, 
\end{array} 
\end{equation} 
which allows us to rewrite \eqref{eq: G matrix} as 
\begin{widetext}
\begin{equation}
\left(\begin{array}{rrrr}
\gamma_{11}&-\gamma_{12}&-\gamma_{13}&-\gamma_{14}\\
-\gamma_{21}&\gamma_{22}&-\gamma_{23}&-\gamma_{24}\\
-\gamma_{31}&-\gamma_{32}&\gamma_{33}&-\gamma_{34}\\
-\gamma_{41}&-\gamma_{42}&-\gamma_{43}&\gamma_{44}
\end{array}\right)
\left(\begin{array}{c}I_{1}\\I_{2}\\I_{3}\\I_{4}\end{array}\right) 
= \Delta V
\left (\begin{array}{r} G_{13}+G_{14}\phantom{)} \\ 
G_{23}+G_{24}\phantom{)} \\ 
-(G_{13}+G_{23}) \\
-(G_{14}+G_{24}) 
\end{array}\right). 
\end{equation} 
Here we have defined $\gamma_{ab}=\delta_{ab}+G_{ab}R_b$ and $\Delta
V=V_L-V_R$. 

Current conservation implies that $\sum I_a = 0$, which allows us to reduce the
problem to a $3\times3$ matrix equation: changing variables to
\begin{equation} 
x = \frac{I_1+I_2}{2},\quad 
y = \frac{I_1-I_2}{2},\quad 
z = \frac{I_3-I_4}{2}, 
\end{equation} 
we have 
\begin{equation} \label{eq:3 by 3 G matrix}
\left(\begin{array}{rrr} 
\gamma_{11} - \gamma_{12} + \gamma_{13} + \gamma_{14} &
\gamma_{11} + \gamma_{12} & -\gamma_{13} + \gamma_{14} \\ 
\gamma_{22} - \gamma_{21} + \gamma_{23} + \gamma_{24} & 
-\gamma_{22} - \gamma_{21} & -\gamma_{23} +\gamma_{24} \\ 
\gamma_{33} + \gamma_{31} + \gamma_{32} -\gamma_{34} & 
\gamma_{31}-\gamma_{32} & -\gamma_{33} - \gamma_{34} 
\end{array}\right) 
\left(\begin{array}{c} x \\ y \\ z
\end{array}\right) = 
\Delta V \left(\begin{array}{c} G_{13}+G_{14} \\
G_{23}+G_{24} \\ G_{31}+G_{32} \end{array}\right).  
\end{equation} 
\end{widetext}

In addition to the time-reversal symmetry mentioned above, we assume a
right-left symmetry with interchange of spin direction. This assumption holds
for the ballistic Hamiltonian of \eqref{eq:hamiltonian}, in which the potential
$\vec{f}(\vec{r})$ is symmetric, but would fail if $\vec{f}(\vec{r})$ were
non-symmetric or if there were an additional non-symmetric potential term (as is
the case when there is disorder in the wall). Restricting ourselves to the
symmetric case, the calculations simplify considerably due to the following
additional equalities: 
\begin{equation}
G_{12} = G_{43},\quad G_{13} = G_{42}.
\end{equation} 
Since $R_1=R_4=R_+$ and $R_2=R_3=R_-$, we also have
\begin{equation}
\begin{array}{lcc} 
\gamma_{11}=\gamma_{44}, & \gamma_{22}=\gamma_{33}, \\
\gamma_{12}=\gamma_{43}, & \gamma_{21}=\gamma_{34}, \\
\gamma_{14}=\gamma_{41}, & \gamma_{23}=\gamma_{32}, \\
\gamma_{13}=\gamma_{42}, & \gamma_{31}=\gamma_{24}.
\end{array} 
\end{equation} 
Substituting these relations into \eqref{eq:3 by 3 G matrix} and subtracting row
3 from row 2 we find $y=z$, from which it follows that $I_{1}=-I_{4}$ and
$I_{2}=-I_{3}$. This is intuitively obvious from the symmetric structure of the
circuit. We are thus left with a $2\times2$ system for $x$ and $y$: 
\begin{widetext}
\begin{equation}\label{eq:2by2}
\left(\begin{array}{rr}
\gamma_{11}-\gamma_{12}+\gamma_{13}+\gamma_{14} & 
\gamma_{11}+\gamma_{12}-\gamma_{13}+\gamma_{14} \\
\gamma_{22}-\gamma_{21}+\gamma_{23}+\gamma_{24} &
-\gamma_{22}-\gamma_{21}-\gamma_{23}+\gamma_{24}
\end{array}\right)
\left(\begin{array}{c}x\\y\end{array}\right) = \Delta V
\left(\begin{array}{c}
G_{13}+G_{14} \\ G_{23}+G_{24}
\end{array}\right).
\end{equation}

The total resistance of the circuit between $V_L$ and $V_R$ is given by
$\RDW=\Delta V/(I_1+I_2)=\Delta V/2x$. Solving \eqref{eq:2by2} for $x$, we
obtain 
\begin{equation} \label{eq:GDW} 
\RDW =
\frac{1+(G_{ut}+G_{uf})R_++(G_{dt}+G_{df})R_-+2(G_{ut}G_{df}+G_{dt}G_{uf})R_+R_-}
{G_{ut}+G_{dt}+(R_++R_-)(G_{ut}G_{df}+G_{dt}G_{uf})}, 
\end{equation} 
\end{widetext}
where 
\begin{equation}\nonumber
\begin{array}{cc} 
G_{ut}=G_{13}+G_{14}, & G_{dt}=G_{23}+G_{24}, \\ 
G_{uf}=G_{12}+G_{14}, & G_{df}=G_{21}+G_{23}.  
\end{array} 
\end{equation} 
Here $G_{ut}$ ($G_{dt}$) represents the total left-to-right conductance for the
incoming spin up (down) channel, while $G_{uf}$ ($G_{df}$) represents the total
conductance with spin flip for the incoming spin up (down) channel.

\eqref{eq:GDW} is valid for arbitrary (symmetric) domain walls. The dependence
on the wall structure is contained in the conductances $G_{ab}$. In the
adiabatic and abrupt limits, these reduce to simple values and our model
reproduces the expected results. In the abrupt limit there is no spin reversal
and hence only $G_{13}$ and $G_{24}$ are non-zero. They can be calculated by
summing the transmission function for each conducting channel across the
interface, which has a simple closed form.\cite{valet1993} In the adiabatic
limit there is complete spin tracking, so that $G_{12}=G_{13}=G_{24}=0$ and only
$G_{14}$ and $G_{23}$ are non-zero. Assuming ballistic transport through the
wall, we then have $G_{14}=(e^2/h)N_\uparrow$ and $G_{23}=(e^2/h)N_\downarrow$.
However, for a wall in the adiabatic limit the assumption $2\lambda\lesssim
l_\phi$ is unlikely to be valid, and it is more reasonable to assume diffusive
transport in each spin channel. We should then take $1/G_{14} =
(2\lambda/l_{sd})R_+$ and $1/G_{23} = (2\lambda/l_{sd})R_-$. 
%The total resistance of the circuit is then the parallel combination of
%$(2+\lambda/l_{sd})R_\pm$, which is identical to the case of a uniformly
%magnetized wire.
 
In the general case, the transport through the wall is intermediate between the
adiabatic and abrupt limits and there is transmission both with and without spin
reversal. In the following section, we calculate $G_{ab}$ for the ballistic
Hamiltonian of \eqref{eq:hamiltonian}, which is a reasonable approximation for
cobalt nanowires of the type used in Ref.~\onlinecite{ebels2000}. For wide walls
(such as those in nickel for example) an approach based on diffusive transport,
as used in Refs.~\onlinecite{levy1997} and \onlinecite{simanek2001}, would be
more appropriate.

\section{Conductance parameters for ballistic wall}

We now discuss the calculation of the coefficients $G_{ab}$ for the particular
case of a ballistic domain wall with small spin splitting. In this case we
assume $l_\pm>\lambda$, so that electrons travelling through the wall experience
no scattering apart from that due to the wall. We assume the magnetic structure
to be one-dimensional, \ie $\vec{f}(\vec{r})\equiv\vec{f}(z)$, so that there is
no scattering between different transverse modes. Micromagnetic simulations
indicate that this assumption is reasonable provided the wire diameter is small
enough ($L_x,L_y\lesssim40\nanometer$ for Co), while for wires of larger
diameter more complicated structures such as vortex walls may be energetically
favourable.\cite{forster2002} 

In Ref.~\onlinecite{gopar2004} it was shown that the electron transport
properties of a domain wall depend on its length and energy scales ($\lambda$
and $\Delta$) but are relatively insensitive to the precise form of the function
$\vec{f}(z)$. For mathematical convenience we assume in this work a
trigonometric profile 
\begin{equation}
\vec{f}(z) = \left\{
\begin{array}{rl}
(\cos{\frac{\pi z}{2\lambda}},0,\sin{\frac{\pi z}{2\lambda}}), &
|z|<\lambda,\\[1.5mm]
(0,0,\sgn{z}), & |z|\ge\lambda.
\end{array} \right.
\end{equation}
The spinor Schr\"odinger equation for the longitudinal wavefunctions with
longitudinal energy $E_z=E_F-E_\perp$ can then be written in dimensionless form
as
\begin{equation}\label{eq:schrodinger} 
\left(\frac{\rmd^2}{\rmd\xi^2}+\epsilon
-\left(\begin{array}{cc}
\sin{\frac{\pi\xi}{2p}} & \cos{\frac{\pi\xi}{2p}} \\ 
\cos{\frac{\pi\xi}{2p}} & -\sin{\frac{\pi\xi}{2p}}
\end{array}\right) 
\right)\Psi(\xi)=0,
\end{equation} 
where $\epsilon=2E_z/\Delta$, $p=\lambda\sqrt{m\Delta/\hbar^2}$ and
$\xi=pz/\lambda$. The dimensionless parameter $p$ characterizes the effective
width of the domain wall, and depends on both the actual width ($\lambda$) and
the spin-splitting energy ($\Delta$). The transmission properties of a
conducting state are completely determined by $\epsilon$, the longitudinal
energy measured relative to $\Delta$.

Scattering state solutions of \eqref{eq:schrodinger} can be found in closed
form, as described in detail in the Appendix of Ref.~\onlinecite{gopar2004}. The
solutions of the trigonometric wall have been used in
Ref.~\onlinecite{brataas1999}, and analytic expressions to first order in
$1/\lambda$ are given in Ref.~\onlinecite{waintal2004}. From these solutions we
obtain transmission and reflection coefficients, $T_{\sigma\sigma'}(\epsilon)$
and $R_{\sigma\sigma'}(\epsilon)$, which give the probability for an electron
incident on the left-hand side of the domain wall in the spin state $\sigma$ to
be transmitted or reflected into the spin state $\sigma'$. Similarly,
$T'_{\sigma\sigma'}(\epsilon)$ and $R'_{\sigma\sigma'}(\epsilon)$ denote the
transmission and reflection for electrons incident from the right. The
conductance matrix elements $G_{ab}$ are obtained by summing the appropriate
transmission or reflection function over all states at the Fermi energy:
\begin{equation} \label{eq:gpq} 
G_{ab}=\frac{e^2}{h}\sum_{n_x,n_y}F_{\sigma\sigma'}(\epsilon).
\end{equation} 
Here $F$ denotes either $T$, $R$, $T'$ or $R'$, depending on the positions of
the terminals $a$ and $b$; $\sigma$ and $\sigma'$ denote the spin
orientation in $a$ and $b$. 

The symmetry requirements considered above for $G_{ab}$ also apply to the
transmission coefficients: from the time-reversal symmetry of
\eqref{eq:schrodinger} we have
$T_{\sigma\sigma'}(\epsilon)=T'_{\sigma'\sigma}(\epsilon)$,
$R_{\ua\da}(\epsilon)=R_{\da\ua}(\epsilon)$ and
$R'_{\ua\da}(\epsilon)=R'_{\da\ua}(\epsilon)$, while from the left-right
symmetry of $\vec{f}(z)$ we have $R_{\ua\da}(\epsilon)=R'_{\da\ua}(\epsilon)$
and $T_{\ua\ua}(\epsilon)=T'_{\da\da}(\epsilon)$. 
%The only functions we need to consider are therefore $R_{\ua\da}(\epsilon)$,
%$T_{\ua\ua}(\epsilon)$, $T_{\ua\da}(\epsilon)$ and $T_{\da\ua}(\epsilon)$. Of
%these, only $T_{\ua\da}(\epsilon)$ is non-zero for energies below the splitting
%height, $-1<\epsilon<1$. 
In Ref.~\onlinecite{gopar2004} the properties of $T_{\sigma\sigma'}(\epsilon)$
and $R_{\sigma\sigma'}(\epsilon)$ as a function of $\epsilon$ and $p$ were
studied in detail. For longitudinal energies $\epsilon>1$, it was found that
$R_{\sigma\sigma'}(\epsilon)\simeq0$ (except for extremely narrow walls) and
hence $T_{\ua\ua}(\epsilon)+T_{\ua\da}(\epsilon)\simeq
T_{\da\ua}(\epsilon)+T_{\da\da}(\epsilon)\simeq1$. Since
$T_{\ua\ua}(\epsilon)=T_{\da\da}(\epsilon)$ by the above relations, this implies
that $T_{\ua\da}(\epsilon)\simeq T_{\da\ua}(\epsilon)$ for $\epsilon>1$. 

In Ref.~\onlinecite{gopar2004} the transmission and reflection functions were
considered for relatively low longitudinal energies. In the present case,
however, we are interested in the transmission over the full range of
longitudinal energies up to $\epsilon_F$, since this is what determines the
conductances $G_{ab}$. In Fig.~3 we plot the functions $T_{\ua\ua}(\epsilon)$
and $T_{\ua\da}(\epsilon)$ for several values of $p$ with $\epsilon$ in the
range 1 to 100. For all $p$, the transmission without spin flip
($T_{\ua\ua}(\epsilon)$) eventually goes to unity as
$\epsilon\rightarrow\infty$, while the transmission with spin flip
($T_{\ua\da}(\epsilon)$) goes to zero. Essentially, this is because states with
large longitudinal energy traverse the wall rapidly and do not spend enough time
in the region of rotating magnetization to undergo a complete spin reversal. As
Fig.~3 shows, the energy range over which $T_{\ua\da}(\epsilon)$ is close to
unity increases with increasing $p$, which means that the transport through the
wall becomes increasingly adiabatic as $p$ increases. The case $p=2.5$ shows
very little spin flip, while $p=20$ is close to the adiabatic limit and shows
almost complete spin flip over the energy range considered. For intermediate
values ($p=5$), there is a co-existence of adiabatic transmission, for states
with low longitudinal energy, and non-adiabatic transmission, for states with
high longitudinal energy.\cite{gopar2004} The oscillatory behaviour evident in
parts ($b$) and ($c$) of Fig.~3 is a consequence of the precessional motion of
the electrons in the wall.

\begin{figure}\label{fig:transmission}
\includegraphics[width=0.45\textwidth]{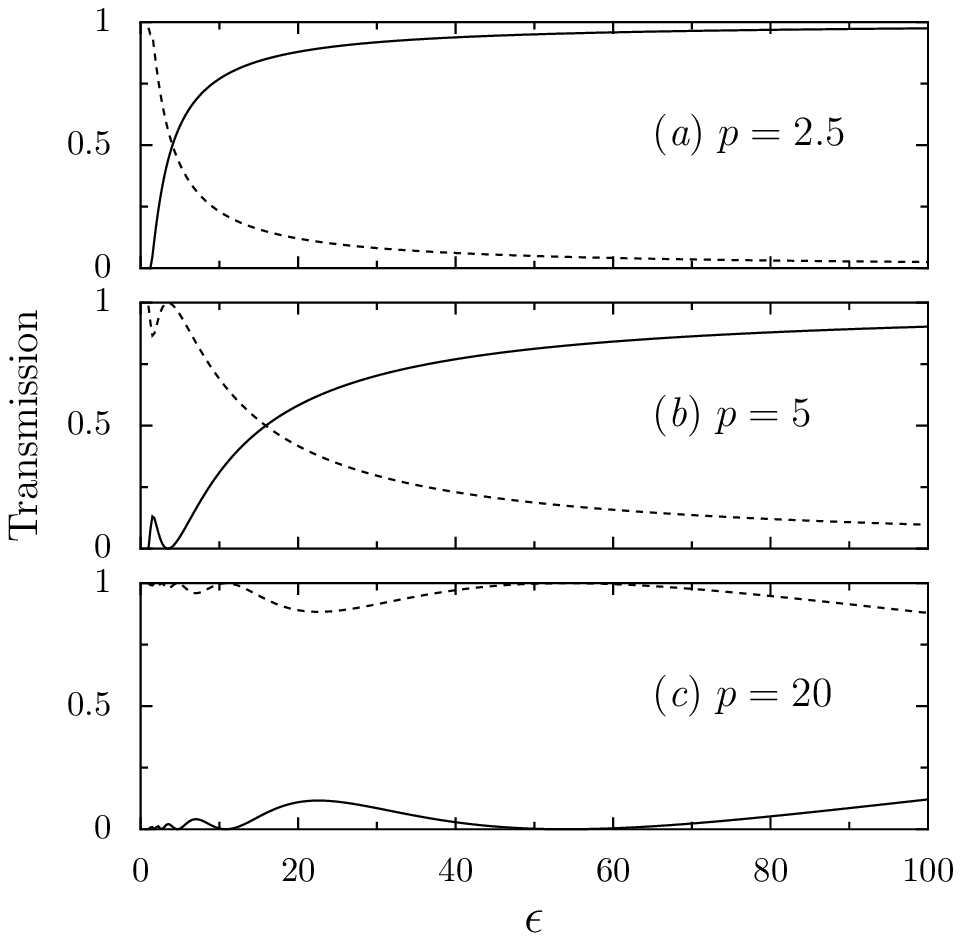}
\caption{Transmission functions $T_{\uparrow\uparrow}(\epsilon)$ (solid lines)
and $T_{\uparrow\downarrow}(\epsilon)$ (dashed lines) as a function of
$\epsilon$ ($=2E_z/\Delta$) for $p$ equal to (\textit{a}) 2.5, (\textit{b}) 5 and
(\textit{c}) 20.}
\end{figure}

The conductances $G_{ab}$ are found using \eqref{eq:gpq}. As mentioned in
Section I, we work in the case of small splitting ($\Delta\ll E_F$), which
allows us to ignore those states which lie below the splitting height
($-1<\epsilon<1$) and consider only states with $\epsilon>1$. As mentioned
above, in this range we have $R_{\ua\da}(\epsilon)\simeq0$, which implies that
$G_{12}\simeq0$, and $T_{\ua\da}(\epsilon)\simeq T_{\da\ua}(\epsilon)$, which
implies that $G_{14}\simeq G_{23}$. Furthermore, the fact that
$T_{\ua\ua}(\epsilon)+T_{\ua\da}(\epsilon)\simeq1$ allows us to write
\begin{equation}\label{eq:P def} 
G_{13}=\frac{e^2}{h}N(1-P),\quad G_{14}=\frac{e^2}{h}NP, 
\end{equation}
where $N\simeq 2mAE_F/\pi\hbar^2$ is the number of channels in the energy range
$1<\epsilon<\epsilon_F$ and we define 
\begin{equation} \label{eq:P integral}
P = \frac{1}{\epsilon_F-1}\int_1^{\epsilon_F}
T_{\ua\da}(\epsilon)\rmd\epsilon. 
\end{equation} 
The parameter $P$ ($0<P<1$) characterizes the amount of conductance with
spin flip, and is the appropriate one to describe transport since it depends not
only on the characteristics of the wall ($p$), but also on the relation between
$E_F$ and $\Delta$ ($\epsilon_F$). In this way, $P$ incorporates the different
degrees of adiabaticity of electrons at the Fermi energy. In Fig.~4 we show $P$
as a function of $p$ for $\epsilon_F=10,100$, calculated using \eqref{eq:P
integral}. For all $\epsilon_F$, there is complete spin reversal (\ie
$P\rightarrow1$) in the limit $p\rightarrow\infty$. However, for smaller
$\epsilon_F$, $P$ goes to $1$ more rapidly since the adiabaticity is most
pronounced for channels with small $\epsilon$. Once again, the precessional
character of the underlying electron motion through the wall gives rise to
subtle oscillations in $P$, which are more pronounced for larger $\epsilon$.

The approximations leading to \eqref{eq:P integral} permit a considerable
simplification of \eqref{eq:GDW}, which can now be written in the form:
\begin{equation}\label{eq:epsilon conductance} 
\RDW =
\frac{r_0(r_0+R_++R_-)+(r_0(R_++R_-)+4R_+R_-)P}{2(r_0+(R_++R_-)P)},
\end{equation} 
where $r_0=h/Ne^2$ is the ballistic resistance of the nanowire. We see that the
total resistance depends only on the spin-dependent resistances ($R_+$ and
$R_-$), the number of channels ($N$) and a single parameter characterizing the
adiabaticity ($P$).

\begin{figure}\label{fig:conductance}
\includegraphics[width=0.45\textwidth]{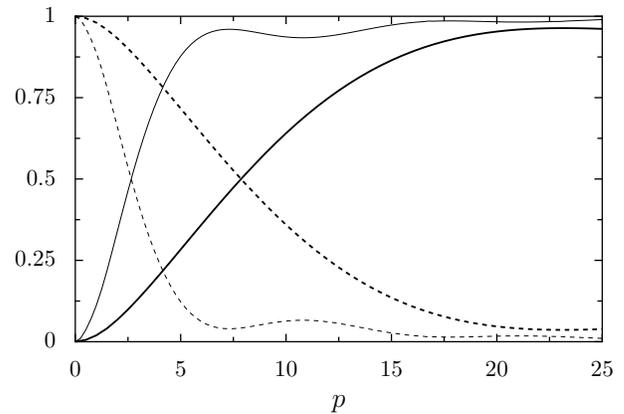}
\caption{Adiabaticity parameter $P$ (solid lines) and $1-P$ (dashed lines) as a
function of the domain wall width parameter $p$ for $\epsilon_F=10$ (thin lines)
and $\epsilon_F=100$ (thick lines).} 
\end{figure}

\section{Domain wall magnetoresistance}

In our circuit model, the presence of a domain wall affects the electronic
transport in a region of length $2l_{sd}+2\lambda$. Letting $\RDW$ and $R_0$
denote the resistance of this region with and without the wall, the (relative)
magnetoresistance due to the wall in a wire of total length $\lwire$ and
resistance $R_\mathrm{wire}$ is
\begin{equation} \label{eq:MR} 
\mathrm{MR} = \frac{\RDW-R_0}{R_\mathrm{wire}}
= \frac{2(l_{sd}+\lambda)}{\lwire}\frac{\RDW-R_0}{R_0}.  
\end{equation}
In the case of the ballistic wall whose conductance parameters we calculated in
the previous section, we can derive a particularly simple and insightful formula
for the magnetoresistance. The resistance of the circuit with domain wall,
$\RDW$, is given by \eqref{eq:epsilon conductance}. The resistance without wall,
$R_0$, is simply the parallel combination of spin up and down resistances, and
can be formally obtained from \eqref{eq:epsilon conductance} by taking the limit
$P\rightarrow1$. Substituting into \eqref{eq:MR} and assuming $r_0\ll R_\pm$, we
find
\begin{equation}\label{eq:dw MR} 
\mathrm{MR} = \frac{l_{sd}+\lambda}{\lwire}\frac{2\beta^2}{1-\beta^2}
\times\frac{1-P}{1+\alpha P}, 
\end{equation} 
where $\alpha=(R_++R_-)/r_0$ and $\beta=(R_--R_+)/(R_++R_-)$. 

\eqref{eq:dw MR} expresses the magnetoresistance as a product of two terms which
depend on a small number of parameters. The first term corresponds to the GMR of
an abrupt interface,\cite{valet1993} and depends on the polarization $\beta$
and the ratio $(l_{sd}+\lambda)/\lwire$. The second term, which is a function of
$P$ and $\alpha$, is a ``reduction factor'' which decreases from 1 to 0 as $P$
goes from 0 to 1. The behaviour of this term is shown in Fig.~5 as a function of
$P$ for several values of $\alpha$. Between the abrupt ($P=0$) and adiabatic
($P=1$) limits, the magnetoresistance decreases monotonically from the full GMR
value to zero. The rate of this transition is determined by $\alpha$: near
$P=0$, where the derivative is $-(1+\alpha)$, the rate of reduction becomes 
steeper with increasing $\alpha$, while at $P=1$ the derivative is equal to
$-1/(1+\alpha)$, and hence in this region the curve becomes flatter with
increasing $\alpha$.

Realistic values of $\alpha$ are reasonably large (on the order of 80 for the
cobalt nanowires in Ref.~\onlinecite{ebels2000}), while the values of $P$ range
from $\sim$$0.25$ for cobalt to $\sim$1 for nickel or permalloy. The region of
Fig.~5 corresponding to realistic physical systems is thus somewhere in the
``tail'' region, in which the magnetoresistance is reduced by approximately an
order of magnitude with respect to the GMR ($P=0$) value. It is, however,
several orders larger than the magnetoresistance predicted by a purely ballistic
model, such as Ref.~\onlinecite{gopar2004}, where the resistance is completely
due to reflection of conduction electrons at the domain wall.

\begin{figure}\label{fig:red factor}
\includegraphics[width=0.45\textwidth]{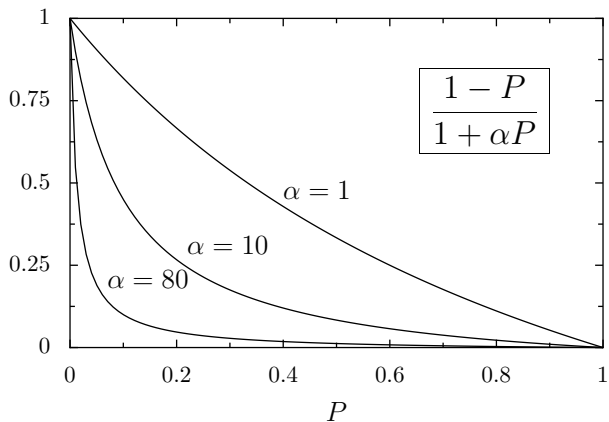}
\caption{The magnetoresistance, as a fraction of the GMR interface value
($P=0$), for $\alpha=1,10$ and $80$.}
\end{figure}

We now compare the predictions of \eqref{eq:dw MR} to experiment. Ebels \etal
\cite{ebels2000} investigated the magnetoresistance due to domain walls in
cobalt nanowires of diameter $35\nanometer$ and found contributions of $0.03\%$
(based on the smallest resistance jump in Fig.~4 of Ref.~\onlinecite{ebels2000})
due to single walls in wires of length $\lwire\simeq20\micrometer$ and total
resistance $R_\mathrm{wire} = 1.4\kohm$.  Using standard material parameters for
cobalt, $l_{sd}=60\nanometer$ and $\beta=0.4$, the authors in
Ref.~\onlinecite{ebels2000} used the interface GMR expression (the first term on
the right-hand side of \eqref{eq:dw MR}) to obtain a magnetoresistance of
$0.14\%$, which is in fact larger than the experimental value.  However, we must
consider the reduction factor in \eqref{eq:dw MR} since the wall has a finite
width. Assuming reasonable parameters for the band structure, $E_F=10\eV$ and
$\Delta=0.1\eV$, and taking a domain wall width equal to the bulk value
$\lambda=10\nanometer$, we find $p\simeq5$, and our ballistic model gives
$P\approx0.25$ and $\alpha\simeq80$. These values lead to a reduction of
$\sim$$0.036$ from the interface value and hence to a magnetoresistance of
$0.0050\%$, which is within an order of magnitude of the experimental result of
Ref.~\onlinecite{ebels2000}.

It is important to note that within a general model such as the one used in this
paper, a quantitative agreement with experiment is not feasible. This is due in
part to the substantial uncertainty in the parameters on which our circuit model
is based: $R_\pm$, $\alpha$, $\beta$ and $P$ (which depends on $p$ and
$\epsilon_F$). Furthermore, in the case of Ref.~\onlinecite{ebels2000}, a direct
comparison is complicated by the fact that the experimental results are strongly
sample-dependent, and there is some uncertainty over the number of domain walls
which contribute to the magnetoresistance measured in the nanowires. Thus, an
agreement to within an order of magnitude should be considered reasonable in
this context.

A related issue is that in nanowires of the type fabricated in
Ref.~\onlinecite{ebels2000}, domain walls tend to be pinned at local
constrictions in which the wire cross-section may be significantly reduced from
its average value. This leads to two effects in our model: firstly, there is an
increase in $r_0$, which is inversely proportional to the number of conducting
channels in the wall region and hence to the cross-sectional area. Secondly, the
presence of a constriction can cause a reduction in the wall width
(characterized by $p$) from the bulk value,\cite{bruno1999} resulting in a
reduction in the adiabaticity, characterized by the parameter $P$. This is
particularly important in the intermediate regime where, as Fig.~4 shows, $P$ is
rather sensitive to changes in $p$. As an example, consider a constriction in
which the diameter and wall width are reduced by a factor of 2 in the
experimental case  of Ref.~\onlinecite{ebels2000}. The reduced wall width
($p\rightarrow2.5$), gives an adiabaticity $P\simeq0.15$ which leads, in
combination with the increased $r_0$, to $\mathrm{MR}\simeq0.03\%$,
corresponding to an increase of an order of magnitude with respect to the
theoretical value without constriction. The presence of a geometric constriction
is thus an important factor which can significantly increase the resistance of a
domain wall.

%There are several other factors which limit the agreement of our model with
%experiment which we mention briefly. Our model assumes a simple spin-split
%parabolic band structure in order to calculate $P$. It has been shown that when
%realistic band structures are taken into account, the results can change
%considerably \cite{vanhoof1999}. We would expect this to play an important role
%in our model as well. For instance, a band structure which featured an enhanced
%number of states with large longitudinal energy would lead to a reduction in
%$P$, since such states are transmitted with a lower amount of spin flip.
%Finally, we have considered explicitly only a ballistic domain wall, which is
%valid for low temperature and narrow walls with a low impurity concentration.
%However, at higher temperature and wide walls, the presence of scattering in the
%domain wall region may be expected to have a significant effect on the spin
%reversal process and hence on the magnetoresistance.  

\section{Spin Transfer Torques and domain wall motion} 

In travelling through a domain wall conduction electrons undergo a reversal of
spin in which angular momentum is exchanged with the wall. If the current
incident on the wall is spin-polarized, there is an overall transfer of angular
momentum from the conduction electrons to the domain wall. This gives rise to a
torque on the domain wall which can cause it to move. This effect was originally
predicted by Berger \cite{berger1984,berger1992} and has recently been observed
in a number of experiments.\cite{grollier2002,vernier2004,yamaguchi2004}

Our circuit model allows us to calculate the total torque exerted on the domain
wall by summing the total spin of incoming and outgoing current components.
Electrons in terminals 1 and 3 carry spin $\hbar/2$, while those in terminals 2
and 4 carry spin $-\hbar/2$. Since $I_4=-I_1$ and $I_3=-I_2$, there is a rate of
angular momentum transfer $(\hbar/e)(I_1-I_2)$ into the domain wall. This is
equivalent to the following torque per unit current:
\begin{equation}\label{eq:torque1}
\frac{\tau}{I}=\frac{\hbar}{e}\,\frac{I_1-I_2}{I_1+I_2}.  
\end{equation} 
In the abrupt (GMR) limit we have $I_1=I_2$ and hence $\tau/I=0$, which is
expected since there is no spin reversal in this case. In the adiabatic limit,
on the other hand, we have $\tau/I=\hbar\beta/e$ ($\beta=(R_--R_+)/(R_++R_-)$),
which corresponds to the result obtained by Berger.\cite{berger1984} For the
intermediate case, \eqref{eq:torque1} yields 
\begin{equation} 
\frac{\tau}{I} =
\frac{\hbar}{e}\,\frac{(G_{ut}-G_{dt})+(R_--R_+)(G_{ut}G_{df}+G_{dt}G_{uf})}
{(G_{ut}+G_{dt})+(R_++R_-)(G_{ut}G_{df}+G_{dt}G_{uf})}.  
\end{equation} 
For a ballistic wall, the results of the previous section yield the simple
expression 
\begin{equation} \label{eq:tau/I}
\frac{\tau}{I}=\frac{\hbar\beta}{e}\,\frac{\alpha P}{1+\alpha P}, 
\end{equation}
where $\alpha$ and $\beta$ are as previously defined. The first part of this
expression, $\hbar\beta/e$, is the adiabatic spin-transfer torque mentioned
above, while the second part is a reduction factor which is less than one when
there is incomplete spin reversal in the domain wall. In contrast to the
magnetoresistance of the previous section, the torque becomes most significant
in the adiabatic limit. We note that in our model, there is always a complete
reversal of spin over the length scale of the circuit (\ie $2l_{sd}+2\lambda$);
\eqref{eq:tau/I} gives the proportion of this which occurs in the domain wall
itself. 

It is important to notice that in a model considering an isolated ballistic wall
(as in Ref.~\onlinecite{gopar2004}) the spin polarization of the current (and
hence the resulting torque) would be proportional to $\Delta/E_F$, and therefore
very small. The inclusion of spin-dependent resistances on both sides of the
wall leads to a much larger difference between up and down current components,
and hence a sizeable torque.

When the precessional nature of the conduction electron motion inside the wall
is taken into account it is found that, in addition to an overall component
perpendicular to the local magnetization direction, there is a spatially varying
torque component.\cite{waintal2004} The former gives rise to motion of the
domain wall, while the latter tends to induce distortions in the wall profile.
However, to a first approximation it is reasonable to assume that the domain
wall is ``rigid'' with respect to the torque exerted by the conduction
electrons, so that only the total torque exerted over the domain wall length is
relevant. 

If the energy of the domain wall is independent of position along the wire, \ie
there is no pinning potential, we can obtain a simple estimate for the velocity
of the wall motion which results from the spin-transfer torque. Suppose the
torque $\tau$ acts on the wall for time $\Delta t$. Then in order to absorb the
transferred angular momentum $\Delta S=\tau\Delta t$, the wall moves by an
amount $\Delta z=\Delta S/\rho_SA$, where $\rho_S$ denotes the angular momentum
density (per unit volume) in the wire and $A$ is the cross-sectional area.
$\rho_S$ can be determined using $\rho_S=M_S\hbar/\mu_B g$, where the saturation
magnetization $M_S$ and gyromagnetic ratio $g$ are material-specific parameters
of the ferromagnet. We thus have a simple formula for wall velocity per unit
current density:
\begin{equation}\label{eq:wall speed} 
\frac{v_{\mathrm{wall}}}{I/A} = \frac{1}{\rho_S}\,\frac{\tau}{I} = 
\frac{\hbar\beta}{\rho_S e}\,\frac{\alpha P}{1+\alpha P}. 
\end{equation}
     
In a recent experiment, Yamaguchi and co-workers measured the displacement of
domain walls under the influence of current pulses of varying duration in
permalloy nanowires of cross-section
$240\nanometer\times10\nanometer$.\cite{yamaguchi2004} Assuming that the domain
wall moves at constant speed, they found an average domain wall velocity of
$3\metre\,\second^{-1}$ for current pulses of density
$1.2\times10^{12}\Amp\,\metre^{-2}$. To compare with \eqref{eq:wall speed}, we
note that for permalloy the domain wall width is approximately
$\lambda\simeq100\nanometer$, which gives an adiabaticity $P\simeq1$, \ie
essentially completely adiabatic. Substituting typical values for the material
parameters, $\beta=0.5$, $M_S=2\,\textrm{Tesla}/\mu_0$ and
$g=2$,\cite{skomski1999} \eqref{eq:wall speed} gives
$v_\mathrm{wall}\simeq300\metre\,\second^{-1}$. This value is two orders of
magnitude larger than experiment, but is in agreement with another recent
calculation.\cite{tatara2004} This suggests that in real physical systems the
efficiency with which the spin angular momentum of the conduction electrons is
transferred into motion of the domain wall is limited by other mechanisms which
are not contained in the present theoretical description. In
Ref.~\onlinecite{tatara2004} it was suggested that generation of spin waves in
the magnetic structure could be one such mechanism.

Finally, we note that in order for the above argument to be consistent, the
constant speed of the wall, $v_{\mathrm{wall}}$, should be incorporated into the
Hamiltonian of the conduction electrons. The Schr\"odinger equation for electron
states in a wire with a wall moving at constant speed can be solved in the same
way as a stationary one simply by changing to the reference frame of the wall.
The longitudinal wavevectors of the states will then be Doppler-reduced due to
the motion of the wall. However, we have just shown that the speed of the wall
is slow ($\sim$$10^2\metre\,\second^{-1}$ for theory,
$\sim$$3\metre\,\second^{-1}$ for experiment) relative to the electron
velocities, which for a Fermi energy of $10\mathrm{eV}$ are on the order of
$10^{6}\metre\,\second^{-1}$. The Doppler-reduction is therefore negligible and
the conduction electron solutions for a moving wall are approximately the same
as for the stationary wall.

\section{Conclusion} 

We have presented a circuit model to describe electron transport through domain
walls in ferromagnetic nanowires. This model is a generalization of the GMR
two-resistor model taking into account the partial reversal of spin experienced
by conduction electrons traversing the wall. In the circuit, the domain wall is
represented as a coherent 4-terminal device connected to classical
spin-dependent resistors. The circuit model is independent of the details of the
transport through the wall and is thus applicable to walls of arbitrary
thickness, in which the transport can be either ballistic or diffusive. 

After deriving a general formula for the resistance of the domain wall circuit,
we considered the case of a ballistic wall. We identified for this case an
appropriate ``adiabaticity parameter'' $P$, representing the average proportion
of reversal of conduction electron spins, which characterizes completely the
transport properties of the wall. Introducing some physically sound assumptions,
we derived a simple formula expressing the magnetoresistance of a domain wall as
a product of two components: a term corresponding to the GMR of an abrupt
interface and a reduction factor taking into account the partial spin reversal
due to the finite width of the wall. The circuit model predicts a
magnetoresistance effect which, although small relative to the interface GMR, is
nevertheless much larger than that predicted by a purely ballistic model, and
within an order of magnitude of the experimental results of
Ref.~\onlinecite{ebels2000}. 

The circuit model also allows a consideration the spin-transfer torque exerted
on the domain wall due to the back-action of the conduction electrons. We
obtained formulas for the torque and wall velocity per unit current, which
predict a value two orders of magnitude larger than a recent
experiment.\cite{yamaguchi2004} This is, nevertheless, in agreement with a
recent theoretical result based on an alternative approach,\cite{tatara2004}
suggesting that there are important physical mechanisms in these experiments
which are beyond the scope of the present theoretical models.

In our model we have considered the most basic elements of the physical system
in order to highlight and study the fundamental and general features of the
problem. In particular, we have assumed a single spin-split parabolic band for
the conduction electrons, which is the simplest non-trivial case in which to
treat conduction as a problem in non-equilibrium transport. Band structure
effects may indeed be significant, but most existing calculations seem to be
predictive only for equilibrium properties, and there are numerous additional
unsolved problems associated with nanostructures in which interfaces strongly
control the underlying atomic scale structure. Calculations taking into account
equilibrium band structures calculated using self consistent local density
approximations \cite{vanhoof1999} have found material-dependent changes in the
resistance of up to two orders of magnitude for purely ballistic models.
Allowing for the importance of band structure and screening effects, our results
have nevertheless shown that the main source of resistance is not scattering
from the domain wall itself, but rather the spin-dependent scattering in the
resistors on either side of the wall. We therefore expect our results to be less
sensitive to the details of band structure and more strongly dependent on
inelastic scattering processes.

For a more realistic treatment of the domain wall, beyond the ballistic case, it
would be desirable to consider the effect of disorder in the domain wall region.
The effect of electron-electron interactions also remains to be investigated. In
the $sd$ model, the most important interaction is between the $s$ and $d$
sub-bands, which we have treated using an effective field. However, this
approach ignores the back-action of the $s$ electrons on the $d$ electrons,
which could give rise to non-trivial dynamics such as spin wave excitation. We
have also ignored many-body effects arising from the electron-electron
interaction within the $s$ band, which has recently been treated in a strictly
one dimensional setting for the case of Luttinger liquids.\cite{pereira2004}
Finally, the effect of variations in the transverse magnetic structure, such as
in vortex walls, could have a significant effect on the transport properties of
the domain wall, especially for wires wider than a few nanometers. In these
cases additional effects on electron momentum due to the vector potential
generated by the domain wall need to be taken into account, as demonstrated by
Cabrera and Falicov.\cite{cabrera1974a}

\begin{acknowledgments} 
PEF is grateful for the support of an Australian Postgraduate Award and a Jean
Rogerson Fellowship from the University of Western Australia and for support
from the Universit\'e Louis Pasteur in Strasbourg. This work was also supported
through the Australian Research Council Linkages and Discovery Programmes
and by the European Union through the RTN programme.
\end{acknowledgments}

\bibliography{bibliography}

\end{document} 
\end